# Time-reversing a monochromatic subwavelength optical focus by optical phase conjugation of multiply-scattered light

**Jongchan Park,[1] Chunghyun Park,[1,2] KyeoReh Lee,[1] Yong-Hoon Cho,[1,2,†] and YongKeun Park[1,*]**

[1]Department of Physics, Korea Advanced Institute of Science and Technology, Daejeon 305-701, Republic of Korea.
[2]KAIST Institute for the NanoCentury, Korea Advanced Institute of Science and Technology, Daejeon 305–701, Republic of Korea.

**Due to its time-reversal nature, optical phase conjugation generates a monochromatic light wave which retraces its propagation paths. Here, we demonstrate the regeneration of a subwavelength optical focus by phase conjugation. Monochromatic light from a subwavelength source is scattered by random nanoparticles, and the scattered light is phase conjugated at the far-field region by coupling its wavefront into a single-mode optical reflector using a spatial light modulator. Then the conjugated beam retraces its propagation paths and forms a refocus on the source at the subwavelength scale. This is the first direct experimental realization of subwavelength focusing beyond the diffraction limit with far-field time reversal in the optical domain.**

A time reversal mirror (TRM) reflects a wave which is a time-reversed solution of the wave equation of the original impinging wavefront in the system. Regardless of the complexity of the system, TRM generates a wave that exactly retraces its original paths, as long as the system satisfies the time-reversal symmetry. This intriguing property of TRMs suggests promising applications, especially for wave transport in inhomogeneous systems where conventional mirrors are unusable. It permits a wave to be delivered through a disordered system without full knowledge of the system. With the aid of developments in electromagnetic devices, TRMs have been demonstrated in acoustic [1] and microwave regimes [2] by directly recording a waveform of interest and generating a time-reversed replication of it.

However, in the optical domain, true means of achieving TRM have not been realized, mainly due to the limited bandwidths of electromagnetic devices, where the oscillation frequencies of optical waves are still several orders faster than the current state-of-the-art technology. Alternatively, a phase conjugation mirror (PCM) can be used to demonstrate the time-reversal of monochromatic light, since the complex conjugation of a monochromatic wave is the time-reversed solution of the wave equation, where the medium is lossless and satisfies the time-reversal symmetry [3]. Historically, PCM was first demonstrated in the 1960s using the holographic principle to record a permanent wavefront on a nonlinear optical material and then replay it with a phase-conjugated reference beam [4]. Later, PCMs were demonstrated with degenerated four-wave mixing methods [5], stimulated Brillouin [6] and Raman [7] scattering methods. Recently, using electromagnetic devices, digital versions of PCMs have been demonstrated in various applications [8-12].

An important aspect of the TRM is its ability to focus beyond the diffraction limit by reversing waves originating from a subwavelength-sized source containing evanescent components [13]. This suggests promising applications in various fields such as telecommunications [14], imaging [15,16], and lithography [17]. In practice, to realize a time-reversal system of waves from a subwavelength source, one also needs to generate a time-reversed replication of the source, such as a subwavelength sink [18], which limits the application potential.

However, recently it was shown that this practical barrier can be overcome by utilizing an inhomogeneous environment [14]. The theoretical analysis states that the time-reversed electromagnetic field from a subwavelength dipole source in an inhomogeneous medium is proportional to the imaginary part of the dyadic Green's function of the system [19] which reflects the spatial distribution of the electromagnetic properties of the surrounding medium [20,21]. After first being realized in the microwave domain [14], there have been attempts to demonstrate this concept in the optical domain. However, although a theoretical study showed promising results [22], it has still not been realized due to the challenging experimental scheme that is required.

Here, we experimentally demonstrate the time-reversal of a subwavelength optical focus containing near-field information, with far-field optical phase conjugation. A monochromatic optical field from a subwavelength light source was scattered by random nanoparticles, which effectively converted the evanescent near-field components of the incident wave to propagating far-field components of the scattered wave. By using a spatial light modulator (SLM), the scattered far-field was coupled into a single-mode optical reflector, which generates a phase-conjugated beam by utilizing its 'single-mode nature' [12]. Due to the reciprocity and time-reversal properties of the multiple light scattering, the phase conjugated beam is refocused into the source position with a subwavelength scale. To the best of our knowledge, this is the first direct experimental realization of generating a subwavelength focus by exploiting the time-reversal nature of waves in the optical domain.

The idea is schematically illustrated in Fig. 1. When a monochromatic light from a subwavelength source is directed onto a conventional mirror, it generates a diverging wavefront following the laws of reflection [Fig. 1(a)]. When the light from the source is reflected by a PCM, the light is redirected into the source position and generates a focus. However, this phase conjugated focus is diffraction limited since the high spatial frequency components of the wave exponentially decay upon propagation [Fig. 1(b)]. In the presence of an inhomogeneous environment where the spatial variation of the dielectric property of the medium is smaller than the size of the diffraction limit, PCM can reverse the subwavelength focus [Fig. 1(c)] by exploiting multiple light scatterings.

To time-reverse the monochromatic optical wave, we employed a digital optical phase conjugation system, which couples the optical field into a single-mode reflector. The wavefront of a scattered field from a disordered medium consisting of random nanoparticles is modulated by a phase-only SLM to couple the optical field into a single-mode reflector consisting of a sub-diffraction limit sized pinhole. Since the two counter-propagating monochromatic waves in a single mode system are in conjugated states, a time-reversed wave is realized by simply reversing the propagation direction of the light at the single-mode pinhole [12].

The use of the single-mode optical reflector avoids the issue that arises from the requirement for precise optical alignment [11], which is extremely challenging, and typically limits the experimental realization of reversing subwavelength focus in the optical domain. Our system does not require precise optical alignment or wavefront correction systems [23] to realize PCM, but as an expense of an iterative optimization process of the SLM.

Figure 2 shows the detailed experimental scheme. A modified commercial near-field scanning optical microscopy (NSOM, WiTec Alpha SNOM) system is used to generate a time-reversed monochromatic subwavelength optical focus together with PCM. A beam from a narrow band diode-pumped solid-state laser ($\lambda$ = 532 nm, 100 mW, Shanghai Dream Laser) is split into a signal arm and a conjugation arm. The signal beam is coupled to an NSOM tip aperture which is in contact with a turbid sample and acts as a subwavelength point source. The light from the subwavelength source is scattered by a turbid medium made up of random zirconium dioxide nanoparticles. A scattered far-field field at the back side of the turbid medium is collected by an objective lens (NA = 0.8, Nikon) and relayed into a PCM by 4-*f* telescopic imaging systems.

The PCM consists of a SLM (x10468-01, Hamamatsu) and a single mode reflector. The SLM was placed at the conjugated plane of the back surface of the sample with a lateral magnification of x600. Thus, the granular size of the far-field scattered speckle pattern imaged on the SLM is about 12 times larger than the pixel size of the SLM. The size of the single-mode pinhole was set to be 20 μm, which is about 1.5 times smaller than the point spread function of the optical system. The intensity signal from the pinhole was collected by a single photon counting module (SPCM, ID100-20, ID Quantique) for optimization of the wavefront, in order to couple the scattered field into the single-mode pinhole. After optimization, a flip mirror was used to couple the conjugation beam to the pinhole from the opposite direction.

We adopted the spatial frequency domain of a transmission matrix as an optimization process; wave vectors were used as an orthogonal basis for optical modes. An optimized phase value for each optical mode was found by measuring intensity signals using SPCM while applying eight different phase values to the wave vector components. To maintain high visibility, the area of the wavefront was divided into two, and each area was optimized individually, while the other area was used as a reference beam. On average, about 240 optical modes were used, and the detected signal at the SPCM was enhanced about 120 times.

After the optimization, the propagating direction of the beam at the pinhole was reversed and the scattered field redirected to the position of the subwavelength light source. This time-reversed near-field signal was acquired by collecting the scattered field from the NSOM tip aperture using a photomultiplier tube. Raster scanning was performed to get a 2D image of the near-field signal at the proximity of the sample.

The experimental result is shown in Fig. 3. A subwavelength point source is generated by the NSOM probe tip [Fig. 3(a)]. The size of the aperture of the tip was measured to be ~55 nm, which defines the size of the subwavelength light source.

At first, the focus was time-reversed in a homogeneous medium; that is, the focus was reversed through the air. A bare glass was placed at the image plane of the SLM. The NSOM probe tip was positioned 10 μm above the bare glass. The light from the NSOM tip was directed to the SLM and optimized. The time-reversal of the optimized beam generated a focus at the position of the NSOM probe tip. The size of the focus was measured by raster scanning of the NSOM probe tip. As shown in Fig. 3(b), the FWHM of the resultant focus was measured to be 780 nm, which is decided by the optical transfer function which the effective numerical aperture of the optical system can support due to diffraction.

Next, we inserted random zirconium dioxide nanoparticles in the proximity of the NSOM tip probe to couple the evanescent near-field component of the light to the propagating far-field. The transmitted far-field from the turbid medium was directed to the SLM and collected by the SPCM. After optimization, the beam was time-reversed by the PCM. The time-reversal of the scattered far-field cause it to retrace its propagation path, and it is focused on the position of the subwavelength source. The image of the resultant focus was acquired by raster scanning of the near-field signal using the NSOM. As expected, the size of the acquired focus was beyond the diffraction limit.

We quantitatively measured the FWHM of the focus [Fig. 3(c)]. The average size of the focus was measured to be 151 ± 17 nm ($\lambda/3.5$, $n = 18$) which is 5 fold smaller than using PCM in the homogeneous environment in our system, and 2 fold smaller than the perfect imaging system of a numerical aperture of 0.8.

In summary, we demonstrated the regeneration of a subwavelength optical focus using random nanoparticles and PCM, which is in principle the time-reversal of a monochromatic light wave. The time-reversal of the scattered far-field rewinds the scrambled paths of the propagating optical waves in a disordered medium, and it couples to evanescent high spatial frequency components to reconstruct a subwavelength optical focus. To the best of our knowledge, this is the first direct demonstration of generating a subwavelength focus with far-field time-reversal in the optical domain.

We acknowledge J. Cho and H. Lee for their technical assistance in sample preparation. This work was supported by KAIST, and the National Research Foundation of Korea (2015R1A3A2066550, 2014K1A3A1A09063027, 2012-M3C1A1-048860, 2014M3C1A3052537, 2016R1A2A1A05005320) and Innopolis foundation (A2015DD126).

[*] yk.park@kaist.ac.kr
[†] yhc@kaist.ac.kr

Reference
[1] M. Fink, D. Cassereau, A. Derode, C. Prada, P. Roux, M. Tanter, J.-l. Thomas, and F. Wu, Rep. Prog. Phys. **63**, 1933 (2000).
[2] G. Lerosey, J. De Rosny, A. Tourin, A. Derode, G. Montaldo, and M. Fink, Physical review letters **92**, 193904 (2004).
[3] R. J. Potton, Rep. Prog. Phys. **67**, 717 (2004).
[4] H. Kogelnik, Bell Syst. Tech. J. **44**, 2451 (1965).
[5] A. Yariv and D. M. Pepper, Opt. Lett. **1**, 16 (1977).
[6] B. Y. Zel'dovich, V. Popovichev, V. Ragul'Skii, and F. Faizullov, in *Landmark Papers On Photorefractive Nonlinear Optics*1995), pp. 303.
[7] B. Y. Zel'dovich and V. Shkunov, Quantum Electronics **7**, 610 (1977).
[8] M. Cui and C. Yang, Opt. Express **18**, 3444 (2010).
[9] Y. M. Wang, B. Judkewitz, C. A. DiMarzio, and C. Yang, Nature communications **3**, 928 (2012).
[10] X. Xu, H. Liu, and L. V. Wang, Nature photonics **5**, 154 (2011).
[11] T. R. Hillman, T. Yamauchi, W. Choi, R. R. Dasari, M. S. Feld, Y. Park, and Z. Yaqoob, Scientific reports **3** (2013).
[12] K. Lee, J. Lee, J.-H. Park, J.-H. Park, and Y. Park, Physical review letters **115**, 153902 (2015).
[13] R. Carminati, J. Saenz, J.-J. Greffet, and M. Nieto-Vesperinas, Physical review A **62**, 012712 (2000).
[14] G. Lerosey, J. De Rosny, A. Tourin, and M. Fink, Science **315**, 1120 (2007).
[15] C. Park *et al.*, Physical review letters **113**, 113901 (2014).
[16] J.-H. Park *et al.*, Nature photonics **7**, 454 (2013).
[17] M. Alkaisi, R. Blaikie, S. McNab, R. Cheung, and D. Cumming, Appl. Phys. Lett. **75**, 3560 (1999).
[18] J. de Rosny and M. Fink, Physical review letters **89**, 124301 (2002).
[19] R. Carminati, R. Pierrat, J. De Rosny, and M. Fink, Opt. Lett. **32**, 3107 (2007).
[20] M. Birowosuto, S. Skipetrov, W. Vos, and A. Mosk, Physical review letters **105**, 013904 (2010).
[21] A. Cazé, R. Pierrat, and R. Carminati, Physical review letters **110**, 063903 (2013).
[22] R. Pierrat, C. Vandenbem, M. Fink, and R. Carminati, Physical Review A **87**, 041801 (2013).
[23] M. Jang, H. Ruan, H. Zhou, B. Judkewitz, and C. Yang, Opt. Express **22**, 14054 (2014).

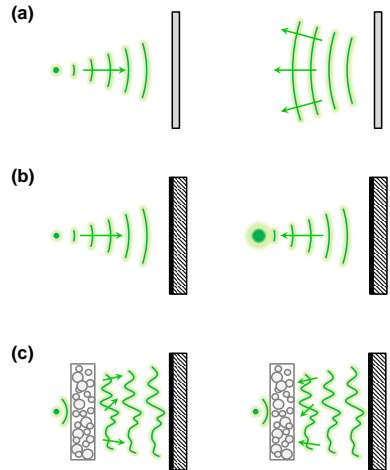

Fig. 1 (color online). Schematic diagram of reversing a sub-wavelength light source. (a) Conventional mirror. (b) Phase-conjugation mirror. (c) Phase-conjugation mirror with random nanoparticles.

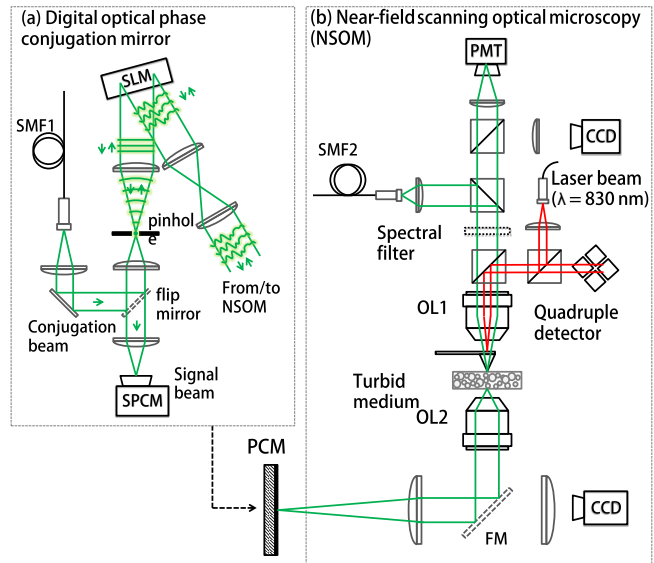

Fig. 2 (color online). Experimental setup. (a) Digital optical phase conjugation mirror. (b) Modified near-field scanning microscopy. SMF: single mode fiber, SPCM: single photon counting module, P: polarizer, SLM: spatial light modulator, PMT: photomultiplier tube, FM: flip mirror, OL: objective lens, CCD: charge-coupled device.

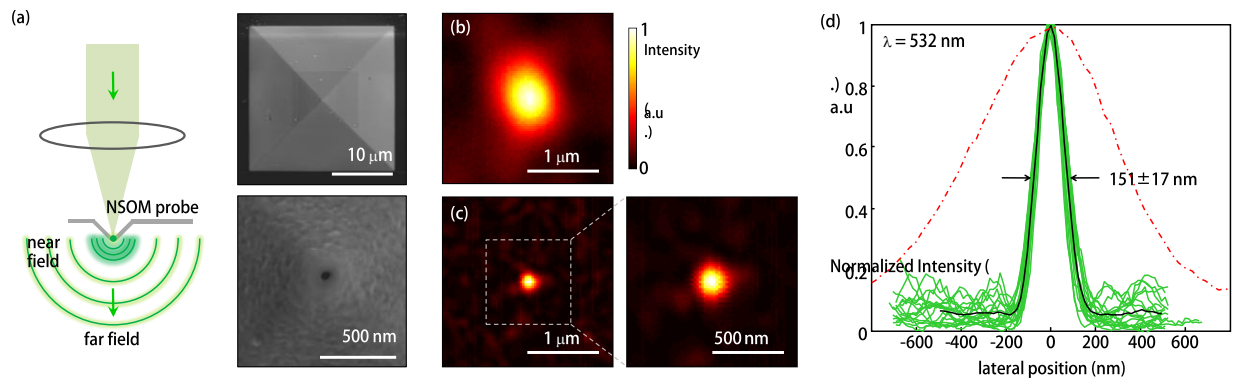

Fig. 3 (color online). (a) Subwavelength focus is made by illuminating NSOM probe tip. Inset, bottom view of the NSOM probe aperture. (b) Far-field time-reversal of the subwavelength focus in homogeneous environment. (c) Using zirconium dioxide random nanoparticles, near-field focus is regenerated by far-field time-reversal of multiply scattered light. (d) Intensity distributions of the foci. Red dash-dotted line, ensemble average of far-field foci. Green solid line, near-field foci (n = 18). Black thick solid line, ensemble average of the near-field foci.